\pdfoutput=1

\documentclass[epj,twocolumn]{webofc}
\usepackage[varg]{txfonts}   
\usepackage{braket}
\usepackage{units}
\usepackage[usenames]{color}

\let\oldleft\left
\def\xleft{\mathopen{}\oldleft}

\newcommand{\bmax}{b_{\rm max}}
\newcommand\bstar{\3{b}_*}
\newcommand\bstarsc{b_*}

\newcommand{\3}[1]{\boldsymbol{#1}}
\newcommand{\T}[1]{\boldsymbol{#1}_{\text{T}}}
\newcommand{\Tsc}[1]{#1_{\text{T}}}
\newcommand{\Tscj}[2]{#1_{#2\,\text{T}}}

\DeclareRobustCommand*\diff[2][]{%
   \mathop{
        \mathrm{d}^{#1}
     \mskip-0.2\thinmuskip
   #2}\nolimits
}

\newcommand{\VC}[1]{%
    \raisebox{-0.5\height}{#1}%
}

\woctitle{TRANSVERSITY 2014}

\begin{document}
\title{Different approaches to TMD Evolution with scale}

\author{John Collins\inst{1}\fnsep\thanks{\email{jcc8@psu.edu}}
}

\institute{%
     104 Davey Lab, Penn State University, University Park PA 16802, USA
}

\abstract{%
   Many apparently contradictory approaches to TMD factorization and
   its non-perturbative content exist.  This talk evaluated the
   different methods and proposed tools for resolving the
   contradictions and experimentally adjudicating the results.
}

\maketitle

\section{Introduction}
\label{intro}

In the literature there is a bewildering variety of methods for using
transverse-momentum-dependent (TMD) parton densities and the
associated factorization properties of cross sections.  Taken at face
value, many of the methods and their uses appear incompatible or
contradictory, especially as regards the non-perturbative
contributions.  The problems are particularly important when planning
new experiments to measure polarization-dependent TMD quantities like
the Sivers function, since the non-perturbative part of TMD evolution
can notably dilute them as energy is increased.

In this talk, I examined and evaluated some of the different methods.
I proposed a systematic approach to test treatments of the
non-perturbative contributions from large transverse distances
($\Tsc{b}$), both from theoretical and phenomenological view points.
Then I proposed systematic modifications to the standard
parameterizations of the large-$\Tsc{b}$ behavior that could resolve
contradictions, especially as regards the apparently incompatible
phenomenology of the function controlling evolution of TMD densities.
The methods will pinpoint the experimental conditions needed to give
incisive experimental probes of the contradictory theoretical
statements.

\section{The need for and existence of non-trivial QCD contributions
  to TMD cross sections}

In this article, I use the Drell-Yan process to illustrate issues
that apply to TMD factorization in general.

For the transverse-momentum distribution in the Drell-Yan process, the
simplest model is the parton model, where the TMD cross section is a
convolution of the TMD densities for the annihilating quark and
antiquark, and the TMD densities do not evolve.  In the parton model,
the transverse momentum of the Drell-Yan pair directly probes the
intrinsic transverse momentum distribution of the quark and antiquark
inside their parent hadrons.

\subsection{Experimental view}

\begin{figure}
\centering
  \newcommand\thisscale{0.41}
  \setlength{\tabcolsep}{4mm}
  \begin{tabular}{cc}
     (a) &
     \VC{\includegraphics[scale=\thisscale]{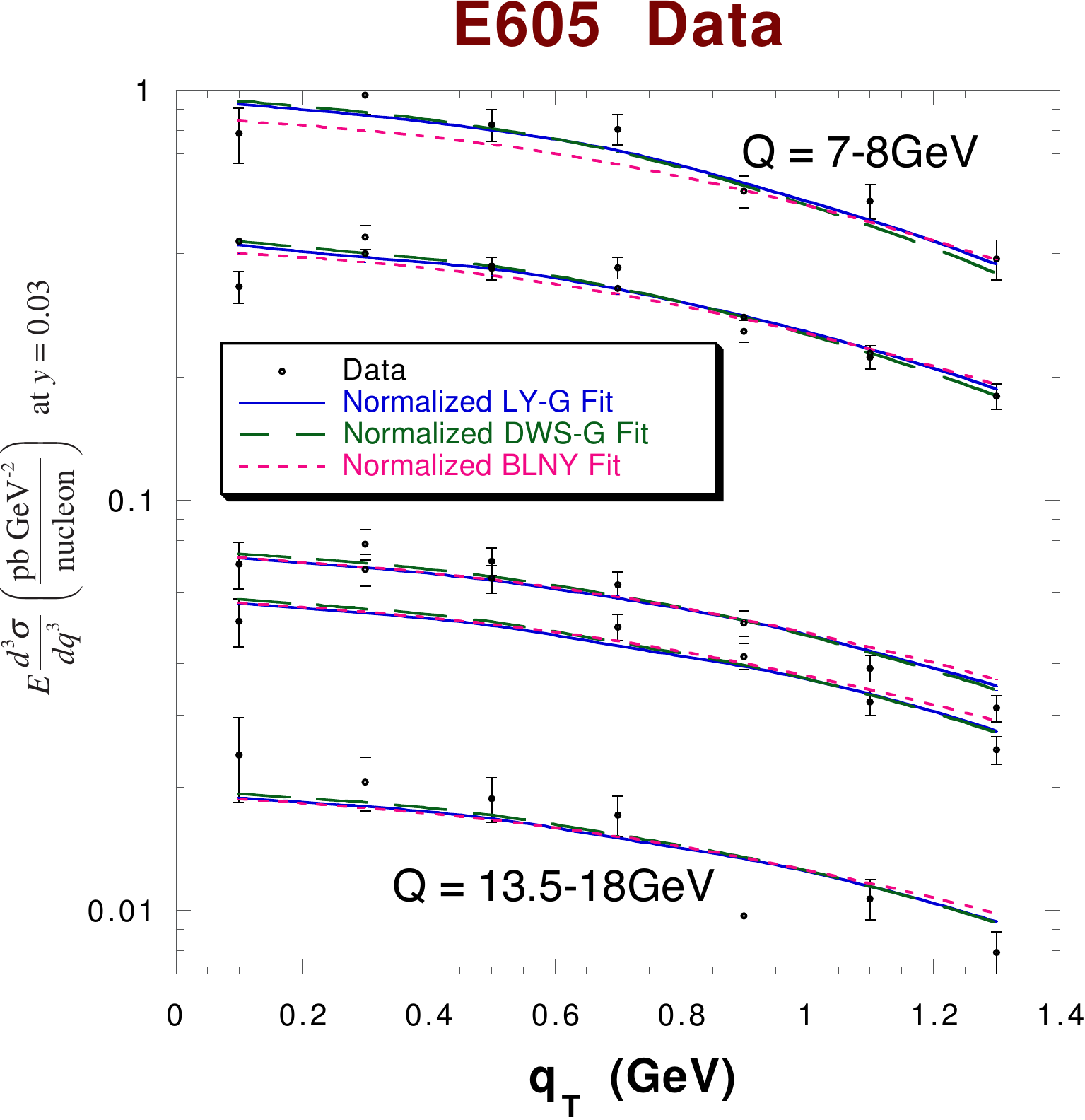}}
  \vspace*{3mm}
  \\
     (b) &
     \VC{\includegraphics[scale=\thisscale]{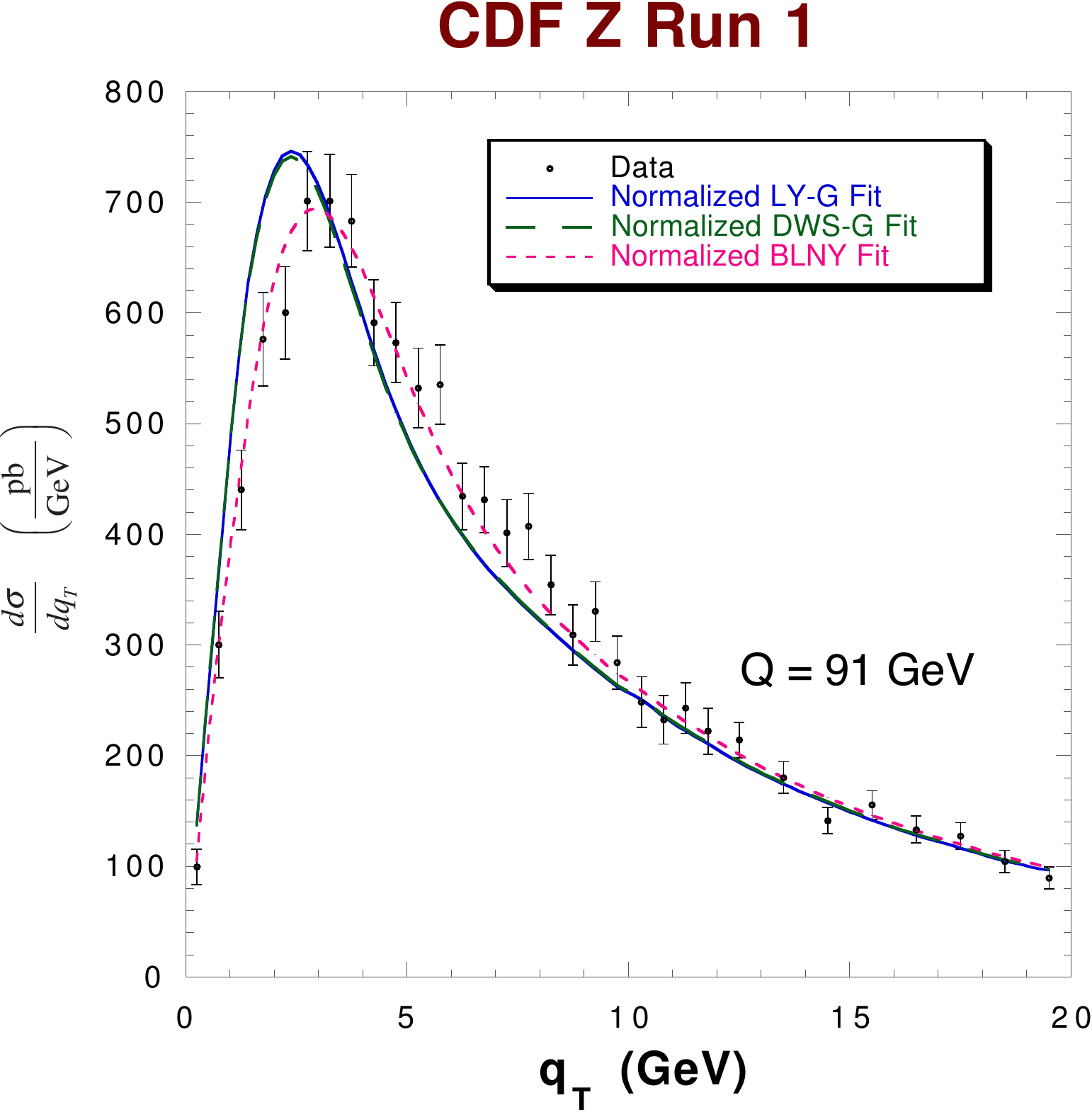}}
  \end{tabular}
\caption{The transverse-momentum distribution in the Drell-Yan process
  at different values of $Q$ and $\sqrt{s}$, showing data from the
  E605 and CDF experiments, together with some fits to the data using
  TMD factorization.
  (Adapted from plots by Landry et al.\ \cite{Landry:2002ix}.)
}
\label{fig:DY.qt}
\end{figure}

That the parton model description is inadequate in reality (and hence
in QCD) is shown by the data in Fig.\ \ref{fig:DY.qt}.  The graphs
also contain several QCD fits to the data.  In plot (a) is shown
$E\diff{\sigma}\!/\!\diff[3]{\3q}$ from the E605 experiment at relatively
low $Q=\unit[7\mbox{--}18]{GeV}$ and $\sqrt{s}=\unit[38.8]{GeV}$.  The
width is around $\unit[1]{GeV}$.

In plot (b) is shown $\diff{\sigma}\!/\!\diff{\Tsc{q}}$ from the CDF
experiment for $Z$ production at $\sqrt{s}=\unit[1800]{GeV}$.  This
has a much larger width, around $\unit[3]{GeV}$.  This value is much
larger than for the lower energy data, and it also appears
incompatible with any reasonable distribution of purely intrinsic
transverse momentum.  It indicates substantial evolution effects, a
specific effect of QCD and other gauge theories.

There is an apparent dramatic difference between the plots at
$\Tsc{q}=0$.  This is merely an artifact of the normalization of the
plotted cross section: Plot (b) has an extra factor of $\Tsc{q}$,
which gives a kinematic zero at the origin; for this plot a sensible
measure of the width of the distribution is the position of the peak.

The values of parton $x$ are characterized by the ratio $Q/\sqrt{s}$,
which is quite different for the two plots.  So interpreting the
difference between the widths as being associated with evolution with
respect to $Q$ is not totally unambiguous; this is recurrent problem.
Actual fits \cite{Landry:2002ix,Konychev:2005iy} use other data as
well, and appear to unambiguously manifest that there is $Q$
dependence at fixed $x$.

\subsection{Need for evolution from QCD}

\begin{figure}
\centering
\setlength{\tabcolsep}{8mm}
\begin{tabular}{cc}
  \VC{\includegraphics[scale=0.35]{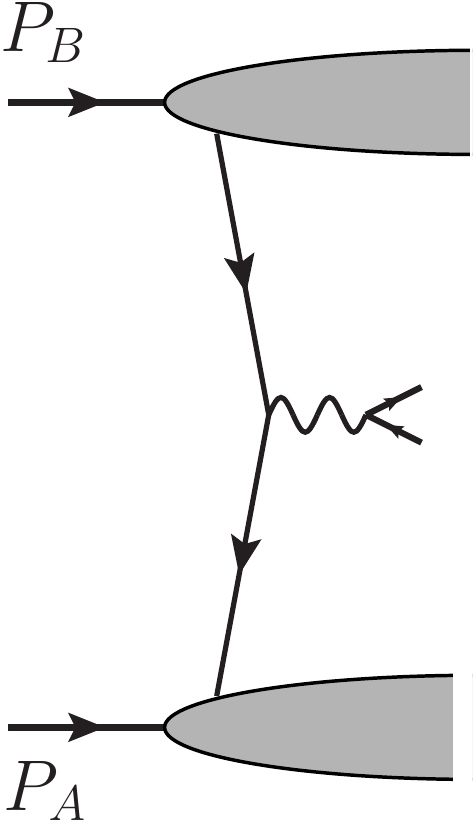}}
 &
  \VC{\includegraphics[clip,scale=0.35]{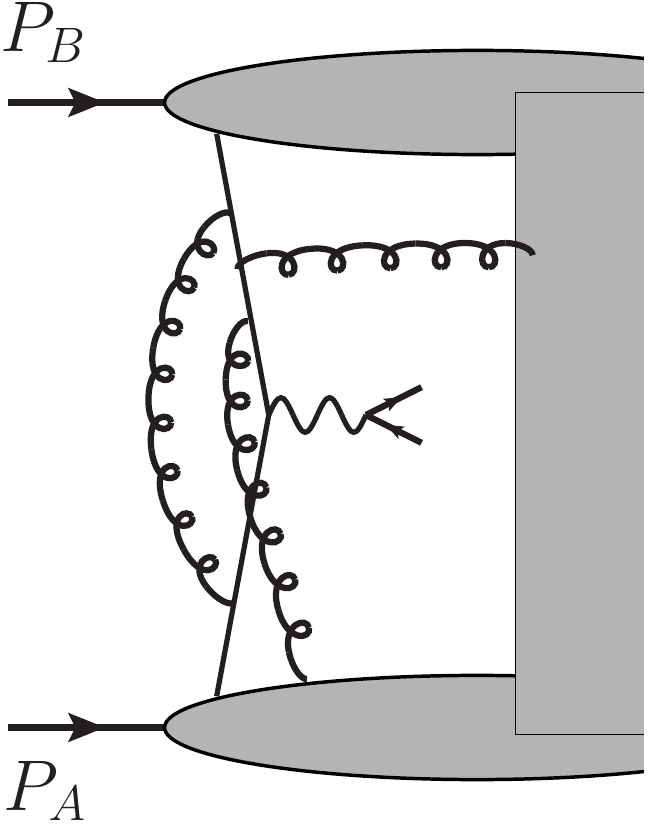}}
\\
  (a) & (b)
\end{tabular}
\caption{For the Drell-Yan process:
   (a) Parton model graphs;
   (b) Examples of leading QCD graphs.
}
\label{fig:DY.graphs}
\end{figure}

That QCD requires substantial modifications to the parton model is
shown on the theoretical side by examining typical graphs that
contribute. In Fig.\ \ref{fig:DY.graphs}(a) is shown the graphical
structure of the amplitude for the Drell-Yan process in the parton
model.  One quark or antiquark out of each of the high-energy incoming
hadrons annihilates to make the Drell-Yan pair; the remaining
``spectator'' parts of the hadrons continue into the final state
unchanged, with a big rapidity gap between them. In the parton model,
other contributions are assumed to be power suppressed.

\begin{figure}
\centering
\begin{tabular}[c]{c}
  \VC{\includegraphics[clip,scale=0.35]{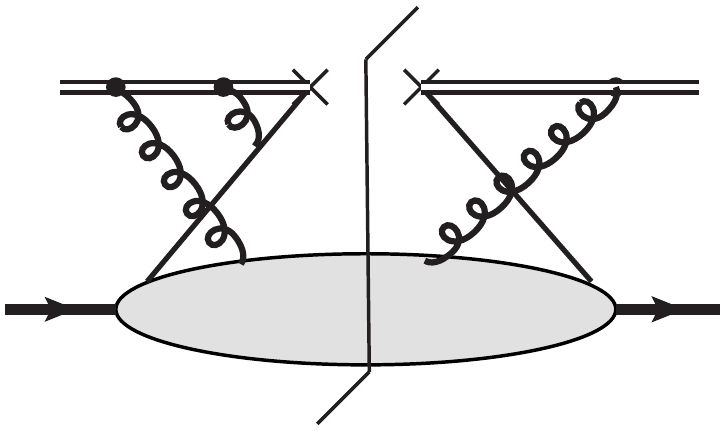}}
\\
  $\text{Fourier trans.\ of } \braket{p|\bar{\psi} ~\text{WL}~  \psi |p}$
\end{tabular}
\caption{Examples of graphs for parton density with Wilson line.}
\label{fig:pdf.graph}
\end{figure}

However, in QCD there are many other contributions that are not
suppressed, as in the example in Fig.\ \ref{fig:DY.graphs}(b).  First,
there are final-state interactions that must exist to neutralize the
color of the spectator parts.  Also, many further contributions exist:
Gluons of any rapidity within the kinematic range set by the incoming
hadrons can connect any of the other lines, including all of: the
active quarks, the spectator parts, and the final-state interaction
component.  Individual graphs do not give a factorized structure.  But
at leading power in $Q$, Ward identities and other methods can be used
to convert the sum over graphs to a factorized form.  The Ward
identities are somewhat unusual, and details can be found in
\cite[Sec.\ 11.9]{Collins:2011qcdbook}.  (Earlier literature is
lacking fully explicit formulations and proofs.)  One consequence is
that the parton densities must be defined with Wilson lines, as in
Fig.\ \ref{fig:pdf.graph}.  Effectively the Ward identities convert
misattached gluons, that link regions of the graph with opposite
rapidities, to attachments to Wilson line operators.  Further
complications involve potential double counting of contributions from
different kinematic regions of internal momenta, which must be
suitably compensated, and the presence of a soft factor that in recent
formulations is absorbed into a redefinition of the TMD densities.

The actual definition of the parton densities is such that the parton
densities have extra scale arguments, and must evolve with energy.
QCD thereby substantially violates the prediction of the pure parton
model that the shape of transverse-momentum distribution scales with
energy.  The broadening arises because gluons are emitted roughly
uniformly into the available range of rapidity, which increases with
energy.  This applies to both perturbative and non-perturbative
gluons.

\section{TMD factorization (modernized Collins-Soper form)}

In this section I summarize the formulae of TMD factorization in the
form I gave in \cite{Collins:2011qcdbook}; detailed proofs were given
there.  Then then I remark on the location of the non-perturbative
information.

\subsection{TMD factorization}

The factorization formula itself for the Drell-Yan cross section is
\begin{align}
\label{eq:fact}
  &\frac{ \diff{\sigma} }{ \diff[4]{q}\diff{\Omega} } 
  = \frac{2}{s} \sum_j
      \frac{ \diff{\hat{\sigma}_{j\bar{\jmath}}}(Q,\mu) }{ \diff{\Omega} }
\times\nonumber\\&\hspace*{5mm}\times
        \int e^{i\T{q}\cdot \T{b} }
         \tilde{f}_{j/A}(x_A,\T{b};\zeta_A,\mu) 
         \tilde{f}_{\bar{\jmath}/B}(x_B,\T{b};\zeta_B,\mu)
        \diff[2]{\T{b}}
\nonumber\\
    & + \mbox{poln.\ terms} 
      + \mbox{high-$\Tsc{q}$ term}
      + \mbox{power-suppressed}.
\end{align}
Here, $\diff{\hat{\sigma}}$ is the hard scattering coefficient, while
the $\tilde{f}_{j/H}(x,\T{b};\zeta,\mu)$ are TMD parton densities
Fourier transformed into transverse coordinate space.  We can set
their scale parameters to $\zeta_A=\zeta_B=Q^2$, $\mu=Q$.

The evolution equations are
\begin{equation}
\label{eq:CSS}
  \frac{ \partial \ln \tilde{f}_{f/H}(x,\Tsc{b}; \zeta; \mu) }
       { \partial \ln \sqrt{\zeta} }
  = 
   \tilde{K}(\Tsc{b};\mu),
\end{equation}
\begin{equation}
\label{eq:RG.K}
  \frac{ \diff{\tilde{K}} }{ \diff{\ln \mu } }
  = -\gamma_K\xleft(\alpha_s(\mu)\right),
\end{equation}
\begin{equation}
\label{eq:RG}
  \frac{ \diff{ \ln \tilde{f}_{f/H}(x,\Tsc{b};\zeta;\mu) }}
       { \diff{\ln \mu} }
    = \gamma_f( \alpha_s(\mu); 1 )
      - \frac12 \gamma_K(\alpha_s(\mu)) \ln \frac{ \zeta }{ \mu^2 }.
\end{equation}
Here, $\tilde{K}(\Tsc{b};\mu)$ is a defined function that controls the
evolution of the TMD pdfs and fragmentation functions of light quarks
with respect to the $\zeta$ parameter.

In the parton model, the integral over all transverse momentum of a
TMD parton density is the corresponding integrated, or collinear
parton density.  Equivalently, when the TMD densities are transformed
to transverse coordinate space, the integrated density equals the TMD
density at zero transverse separation.  In any renormalizable quantum
field theory, this result generally needs to be modified.  Instead,
there is a kind of operator-product expansion (OPE) that expresses the
TMD density at small $\Tsc{b}$ in terms of the integrated densities:
\begin{multline}
\label{eq:OPE}
  \tilde{f}_{f/H}(x,\Tsc{b};\zeta;\mu) 
  = \sum_j \int_{x-}^{1+}
       \tilde{C}_{f/j}\xleft( x/\hat{x},\Tsc{b};\zeta,\mu,\alpha_s(\mu) \right)
\times\\\times
       f_{j/H}(\hat{x};\mu)
       \frac{ \diff{\hat{x}} }{ \hat{x} }
~+~ O\xleft[(m\Tsc{b})^p \right].
\end{multline}
The coefficients are perturbatively calculable provided that the TMD
densities are evolved to scales that avoid large logarithms.  The
lowest-order value of the coefficients is
$\delta_{jf}\delta(x/\hat{x}-1)$, which is the parton model result.

\subsection{Location of non-perturbative information}

The TMD-specific non-perturbative information is at large-$\Tsc{b}$.
Given the existence of the evolution equations, the necessary
information is
\begin{itemize}
\item In the parton densities at large $\Tsc{b}$
  $\tilde{f}_{j/A}(x_A,\T{b};\zeta_A,\mu)$ at one particular scale.
  One may choose to label this the ``intrinsic transverse momentum''
  distribution if the scale is low, although this terminology is not
  entirely accurate.
\item In the evolution kernel $\tilde{K}(\Tsc{b};\mu)$ at large
  $\Tsc{b}$.  This gives a universal character to the evolution, and
  can be characterized as giving the effect of ``soft glue per unit
  rapidity''. 
\end{itemize}
Predictions for cross sections can only be made with the aid of
phenomenological fits for these functions, and/or with the aid of
non-perturbative theoretical modeling and calculation.  The
predictive power of the formalism stems from the universality of these
functions: they can be measured from a limited set of data and used to
predict cross sections in many other situations, with the aid of
evolution and of perturbative calculations of the remaining quantities
needed. 

The OPE at small-$\Tsc{b}$ also needs the values of the ordinary
integrated parton densities.  These are obtained from fits to other
data than is relevant for TMD factorization.  This part of the
non-perturbative information is therefore the same as in collinear
factorization.

\section{Formalisms used}

A list of some of the formalisms that have been used in recent years
is:
\begin{description}
\item[Parton model:] Here QCD complications, especially TMD evolution,
  are ignored.
\item[Non-TMD formalisms:] These eschew the use of TMD densities in
  favor of collinear factorization and a resummation of large
  logarithms in the massless hard scattering.  An old example is by
  Altarelli et al.\ \cite{Altarelli:1984pt}; a recent one is by Bozzi
  et al.\ \cite{Bozzi:2005wk}.
\item[Original CSS:]  Here a non-light-like axial gauge was used to
  define TMD densities without Wilson lines, and a soft factor
  appeared in the TMD factorization formula.
\item[Ji--Ma--Yuan \cite{Ji:2004wu}:]  They implemented the CSS method
  with gauge-invariant TMD densities with non-light-like Wilson
  lines.  They still had a soft factor, and used another parameter
  $\rho$ beyond the scale parameters of CSS.
\item[New CSS:] Here \cite{Collins:2011qcdbook} there is a clean up
  relative to the original CSS version, Wilson lines are mostly
  light-like, and (square roots of) the soft factor are absorbed into
  TMD densities, in such a way that rapidity divergences associated
  with light-like Wilson lines cancel.
\item[Becher--Neubert (BN) \cite{Becher:2010tm}:] This work uses SCET.
  TMD parton densities appear, but they are never finite.
\item[Echevarr\'{\i}a--Idilbi--Scimemi \cite{Echevarria:2012pw}:] This
  is a SCET-based formalism, but with a different regulator to handle
  the divergences given by light-like Wilson lines than is used in the
  CSS and BN formalisms.
\item[Mantry--Petriello \cite{Mantry:2009qz, Mantry:2010bi}:] Another
  SCET-based method. 
\item[Boer \cite{Boer:2008fr}, Sun-Yuan \cite{Sun:2013dya,
    Sun:2013hua}:] These authors start from the CSS formalism, but
  make certain approximations.  Sun and Yuan use no non-perturbative
  function for TMD evolution.
\end{description}

There is disagreement on size of non-perturbative contribution to
evolution, i.e., on the form at large $\Tsc{b}$ of the function that
CSS call $\tilde{K}(\Tsc{b})$; there is even disagreement as to
whether this non-perturbative contribution exists.

\section{Examination of some of the methods}

\subsection{Parton Model}

The factorization formula (\ref{eq:fact}) reduces to the parton model
formula when the hard scattering is replaced by its lowest-order
approximation, TMD evolution is ignored, and the high-$\Tsc{q}$
correction term is ignored.  The parton-model approximation is
typically used to fit data at relatively low energies compared with
the earlier Drell-Yan fits.  At these energies, a particular interest
is in fitting polarization-dependent functions like the Sivers and
Collins functions, e.g., \cite{Anselmino:2013vqa,Anselmino:2011gs}.
Typically a Gaussian ansatz is used for the shape of the TMD
functions, e.g., \cite{Anselmino:2013vqa}.

The OPE (\ref{eq:OPE}) for the TMD densities at small $\Tsc{b}$ shows
that the Gaussian ansatz cannot be exactly correct and that the
Gaussian ansatz will fail once large enough transverse momenta are
considered.  But it evidently allows a good fit to data at low
energy. 

The neglect of higher-order terms in the hard scattering is
reasonable, since $\alpha_s(Q)$ is small.  It is also reasonable to neglect
the high-$\Tsc{q}$ correction when $\Tsc{q}$ is small enough compared
with $Q$.  However, in view of the TMD evolution effects definitely
seen at high $Q$, omitting evolution is not correct when a broad
enough range of $Q$ is considered.

However, in reality it is found \cite{Anselmino:2012re,Aidala:2014hva}
that the data indicates that between the energies of the HERMES and
COMPASS experiments, TMD evolution appears to exist but is weak.  A
complication in coming to this conclusion is that in an experiment at
fixed energy, $x$ and $Q$ are highly correlated.

\subsection{Methods without TMD functions}

Some authors, e.g., Altarelli et al.\ \cite{Altarelli:1984pt} and
Bozzi et al.\ \cite{Bozzi:2005wk}, eschew completely the use of TMD
densities.  They use collinear factorization together with a
resummation of large logarithms of $Q/\Tsc{q}$ in higher orders of the
massless hard scattering coefficient in the collinear factorization
framework.  If this were fully justified, it would improve predictive
power, since the only non-perturbative information used is in the
ordinary integrated parton densities.

However, the justification of collinear factorization uses
approximations for large $Q$ that are valid only when $\Tsc{q}$ is of
order $Q$ or when $\Tsc{q}$ integrated over.  The logical foundation
fails when $\Tsc{q}\ll Q$.  The errors in collinear factorization
relative to the true cross section are suppressed by powers not only
of $\Lambda/Q$ but also of $\Tsc{q}/Q$.  An important symptom of this is
that in the leading power ``twist-2'' collinear factorization, the
effects of Boer-Mulders and Sivers functions are missed, whereas at
low transverse momentum these functions given leading power effects.
See Ref.\ \cite{Bacchetta:2008xw} for a good description of this last
issue. 

Further, in the resummation formalism, integrals over scale include
non-perturbative regions with, e.g., $\alpha_s(k^2)$ at small $k$.  A
proper TMD factorization shows what to in this region.

\subsection{Original CSS}

In the original CSS formalism \cite{Collins:1981uk,Collins:1981uw},
TMD parton densities were defined in a non-gauge-invariant way with
use of non-light-like axial gauge; this was used to cut off the
rapidity divergences that would appear if the most natural definition,
with light-cone gauge, were used.  The CSS evolution formula, of the
form of (\ref{eq:CSS}), gave the dependence of TMD functions on this
rapidity cut off.  There was a separate soft function in the
factorization formula.  Furthermore the evolution equations have
power-suppressed corrections, which are dropped in phenomenological
applications.  

CSS recognized that there are non-perturbative effects at large
transverse distance $\Tsc{b}$.  To separate these from perturbatively
calculable phenomena, they proposed \cite{Collins:1984kg} their
$\bstarsc$ prescription.  The combination of TMD factorization, TMD
evolution and the definitions of the TMD densities etc determined what
kinds of functions to use for parameterization of non-perturbative
parts of the cross section.

Phenomenologically, classic fits to Drell-Yan with $\unit[5]{GeV}\lesssim Q \leq
m_Z$ were made by Landry et al.\ (BLNY) \cite{Landry:2002ix}, and
later by Konychev and Nadolsky (KN) \cite{Konychev:2005iy}.

On the theoretical side, a difficulty with the use of axial gauge to
define parton densities is that the singularities in gluon propagators
prevent the direct use of the contour deformations that are used in
showing that the effects of the Glauber region cancel in the inclusive
Drell-Yan cross section.  CSS did not present an explicit solution to
this problem.  Nevertheless the structure of the formula they
presented for the solution of the evolution equations remains as an
actually implemented method for comparison with data, and agrees with
later results.

\subsection{Ji-Ma-Yuan}

Ji, Ma and Yuan \cite{Ji:2004wu} converted the CSS formalism so that
the TMD densities were defined gauge-invariantly, with non-light-like
Wilson lines.  Their factorization formula still has a separate soft
factor, like that of CSS.  The way in which they derived factorization
entail the use of an extra (dimensionless) $\rho$ parameter in the hard
scattering etc, with $\rho$ being large.  There are associated large
logarithms, and the $\rho$ parameter is in addition to the scale
parameters of the CSS formalism.  There should have been evolution
equation for $\rho$, but such an equation appears not to have been
given.

I know of no fits that actually use this scheme.  Fits continued to
use the CSS method.

\subsection{New CSS}

In \cite{Collins:2011qcdbook}, I derived an updated, improved version
of the CSS results.  On the theoretical side:
\begin{itemize}
\item Covariant gauge was used throughout, with suitable Wilson lines
  in gauge-invariant definitions of all the TMD functions.
\item Full proofs (at least to all orders of perturbation theory) were
  given, including a proof of cancellation of the effects of the
  Glauber region that applies both to collinear and to TMD
  factorization.  (This entails particular directions for the Wilson
  lines.) 
\item A square root of the soft factor was absorbed into each TMD
  parton density and fragmentation functions (in a rather unexpected,
  but unique way).
\item As many Wilson lines were made light-like as possible.  The
  limits are quite non-trivial to formulate, which is a problem that
  stymied Ji, Ma and Yuan.
\item The evolution equations are strictly homogeneous.
\end{itemize}
The result is substantially cleaner methods relative to the original
CSS work.  From a phenomenological viewpoint, the new results should
be regarded as being at most a scheme change from the original CSS
method, as represented by the solution of the evolution equations.

\subsection{Becher-Neubert}

Becher and Neubert \cite{Becher:2010tm} obtained a kind of TMD
factorization in the framework of soft-collinear effective theory
(SCET) in the Beneke-Smirnov style.  The results are intended to be
valid for large $Q$ with $\Tsc{q} \ll Q$, but with a restriction to
$\Tsc{q}\gg\Lambda$ (unlike the CSS framework, which does not have
this last restriction).  By the restriction to $\Tsc{q}\gg\Lambda$,
they evade issues of a full TMD formalism and the need for
non-perturbative information at large $\Tsc{b}$.  But this also means
that their method does not apply in the region of low $\Tsc{q}$, which
is of much experimental interest. Thus their methods also do not
include the physics associated with Sivers and Boer-Mulders functions,
etc, which at leading power show their characteristic effects
primarily in the region of non-perturbative $\Tsc{q}$.

Furthermore they could not define separate TMD pdfs; only the product
of two TMD pdfs was defined and free of divergences.  This represents
an inadequacy of the Beneke-Smirnov approach. 

However, the Becher-Neubert method has given an important tool for
NNLO calculations of the coefficient functions in the OPE
(\ref{eq:OPE}) --- see \cite{Gehrmann:2012ze,Gehrmann:2014yya}.

\subsection{Echevarr\'{\i}a--Idilbi--Scimemi}

Echevarr\'{\i}a, Idilbi and Scimemi \cite{Echevarria:2012pw} also
obtained TMD factorization in a SCET framework.  Their methods are
characterized by the use of strictly light-like Wilson lines, but with
a different kind of regulator for the associated rapidity divergences.
(I do not think it obeys gauge invariance, which causes considerable
difficulty in constructing full proofs.  Full proofs of factorization
make essential use of Ward identities or some equivalent to combine
and cancel non-factorizing terms from individual graphs.)

As with the method of \cite{Collins:2011qcdbook}, they absorb soft
factors into the definition of TMD parton densities, but in a simpler
way that depends on their methodology.  Individual TMD parton
densities are defined, unlike the case for Becher and Neubert's
approach. 

In phenomenological fits, Gaussian parameterizations are used for the
TMD parton densities at an initial scale. But a claim is made that
non-perturbative information is not needed in their equivalent of
CSS's $\tilde{K}$ function that controls the evolution of the shape of
TMD functions.  Instead, for $\tilde{K}$, they use a resummation of
perturbation theory.  This is applied up to a scale of
$\Tsc{b}=\unit[4]{GeV^{-1}}=\unit[0.8]{fm}$ or beyond.

\begin{figure}
\centering
    \VC{\includegraphics[scale=0.7]{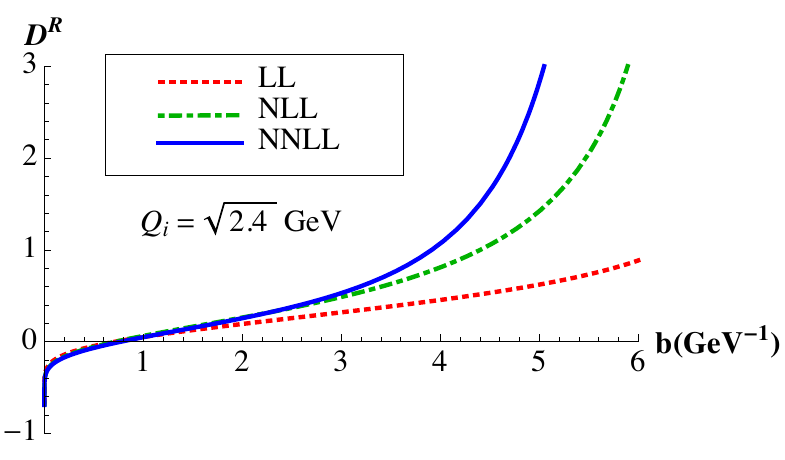}}
    \caption{Plot of $D^R(\Tsc{b};Q_i)=-\tilde{K}(\Tsc{b};Q_i)$, from
      Melis, QCD Evolution 2014 workshop.  Numerical results of three
      approximations are shown: leading logarithm (LL),
      next-to-leading-logarithm (NLL), and
      next-to-next-to-leading-logarithm (NNLL).}
\label{fig:EIS}
\end{figure}

In Fig.\ \ref{fig:EIS} is shown an example of their results for
$\tilde{K}$, in various approximations.

\section{Geography of evolution of cross section}

\begin{figure}
\centering
  \newcommand\thisscale{0.35}
  \begin{tabular}{cc}
     (a) &
     \VC{\includegraphics[scale=\thisscale]{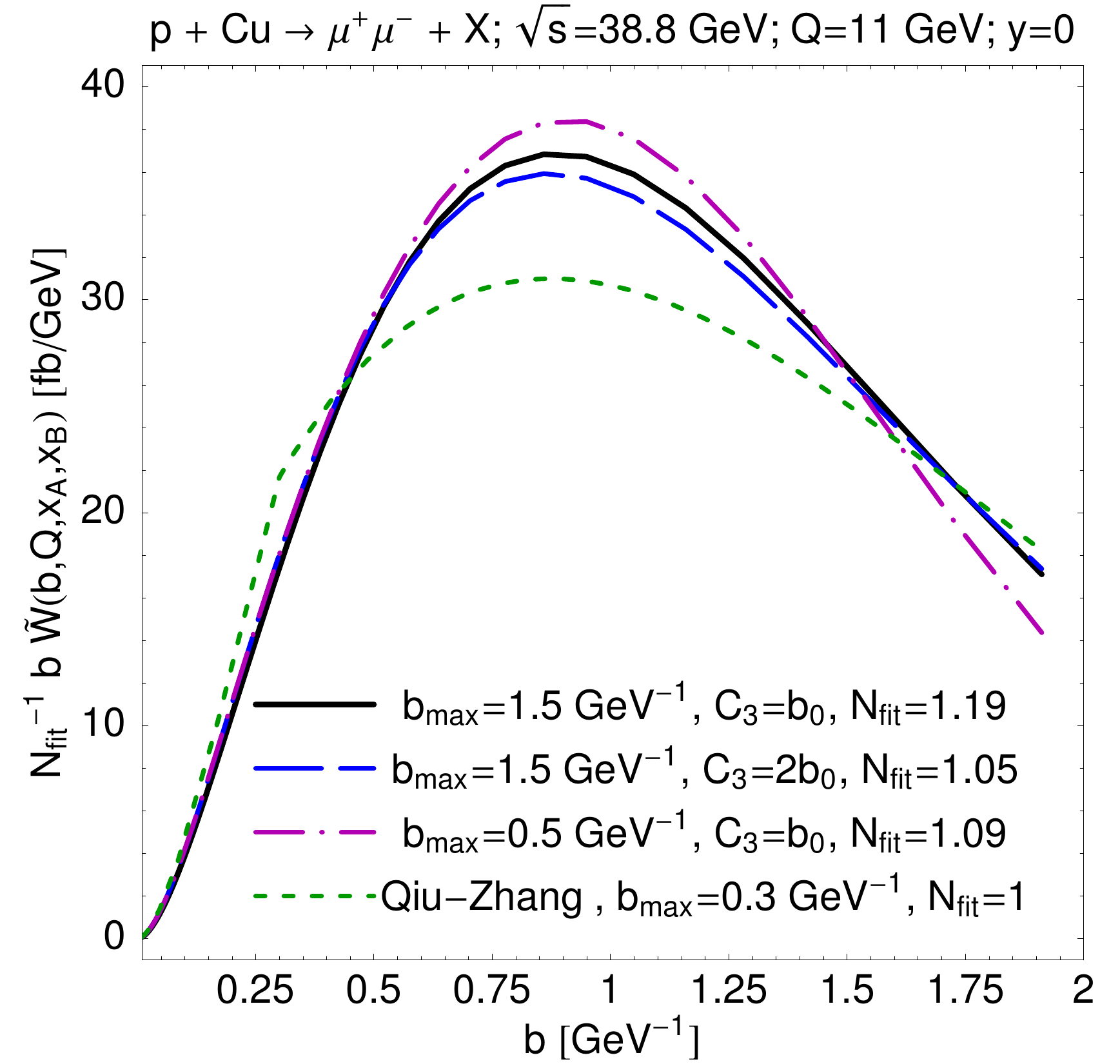}}
  \vspace*{3mm}
  \\
     (b) &
     \VC{\includegraphics[scale=\thisscale]{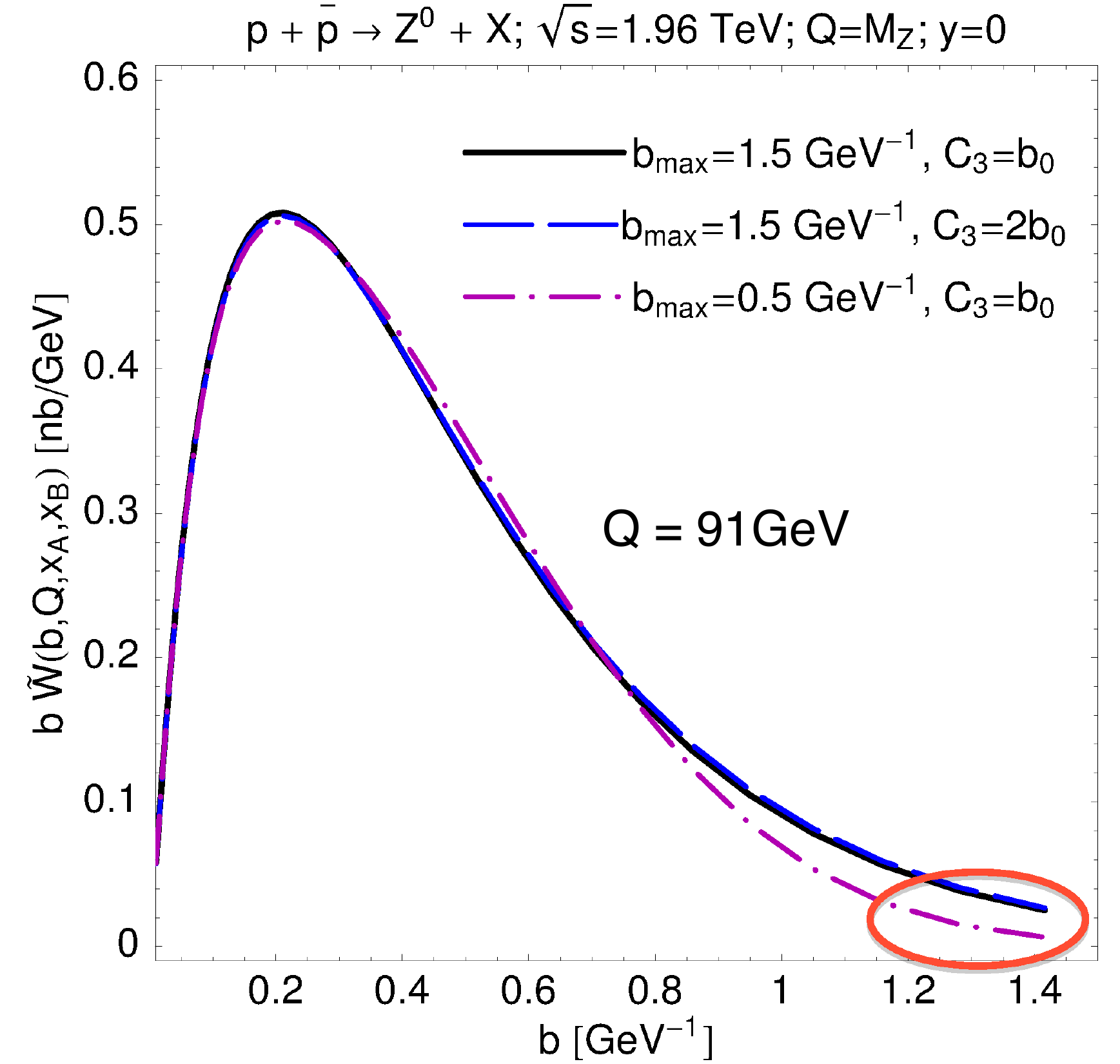}}
  \end{tabular}
  \caption{Plots of the $\Tsc{b}$-space integrands corresponding to
    the cross section plots in Fig.\ \ref{fig:DY.qt}. Adapted from
    plots by Konychev and Nadolsky \cite{Konychev:2005iy}.
  }
\label{fig:evol}
\end{figure}

The evolution of TMD parton densities in formulated multiplicatively
in the space of transverse position.  In Fig.\ \ref{fig:evol} is
plotted the $\Tsc{b}$-space integrand corresponding to the two cross
section plots in Fig.\ \ref{fig:DY.qt}. Up to an overall normalization
factor, the integrand plotted is $\Tsc{b}$ times the integrand in the
TMD factorization formula (\ref{eq:fact}) when $\mu=Q$.  To get the
cross section, this integrand is to be multiplied by the Bessel
function $J_0(\Tsc{q}\Tsc{b})$ and integrated over $\Tsc{b}$ from zero
to infinity.  In general, in going from low to high $Q$, the peak
region of the integrand migrates to ever-smaller values of $\Tsc{b}$.

We now examine the plots with a black solid line and a purple
dot-dashed line.  These correspond to fits made to the same data by
Konychev and Nadolsky \cite{Konychev:2005iy} with the same theoretical
conditions except that $\bmax=\unit[1.5]{GeV^{-1}}=\unit[0.3]{fm}$ and
$\bmax=\unit[0.5]{GeV^{-1}}=\unit[0.1]{fm}$, respectively, for the two
lines.  At lower energies, in graph (a), the two plots do not differ
greatly.  At high energy, in graph (b), the two lines match even more
closely up to about $\Tsc{b}=\unit[0.8]{GeV^{-1}}$, and then they
diverge strikingly, so that the line corresponding to the smaller
value of $\bmax$ is a factor of about two below the other line at the
right-hand edge of the graph.  Although this is a large difference, it
occurs in a region where the integrand is small, so that the large
difference has little effect on the actual cross section.  The
calculation of the cross section is dominated by much smaller values
of $\Tsc{b}$, which are in a perturbative region.

In both cases, the non-perturbative part of $\tilde{K}(\Tsc{b})$ was
parameterized by a quadratic function of $\Tsc{b}$, but the
coefficient is substantially larger for the fit with the small value
of $\bmax=\unit[0.5]{GeV^{-1}}$.  The plot illustrates a general
phenomenon.  Although the integral to get the cross section needs an
integral over all $\Tsc{b}$, up to $\infty$, there is little sensitivity at
large $Q$ to the detailed properties of the integrand at large
$\Tsc{b}$, and hence little sensitivity to the non-perturbative
dependence at large $\Tsc{b}$.

\section{Standard fits of TMD evolution give bad low-$Q$ predictions}

The standard fits (to Drell-Yan data at $Q$ from $\unit[5]{GeV}$ to
$m_Z$) use a quadratic form for $\tilde{K}$, $\tilde{K}(\Tsc{b},\mu)
\propto -\Tsc{b}^2$, at large $\Tsc{b}$.  When the TMD pdfs are
evolved backwards, to lower $Q$, this results in unphysical behavior.
To see this, consider the large-$\Tsc{b}$ behavior of the integrand
for the cross section, as given in:
\begin{equation}
\label{eq:large.b}
  \int \diff[2]{\T{b}} e^{i\T{q}\cdot \T{b} }
      e^{ - \Tsc{b}^2 [\text{coeff}(x) + \text{const.} \times \ln (Q^2/Q_0^2)] }
      \dots.
\end{equation}
The $x$-dependent coefficient is to be obtained from a standard
Gaussian fit to data of TMD densities at some initial scale.  The
coefficient with the $\ln(Q^2/Q_0^2)$ factor in the exponent results
from applying the CSS equation (\ref{eq:CSS}) with a quadratic fit for
$\tilde{K}(\Tsc{b})$ at large $\Tsc{b}$.

\begin{figure}
\centering
    \includegraphics[scale=0.7]{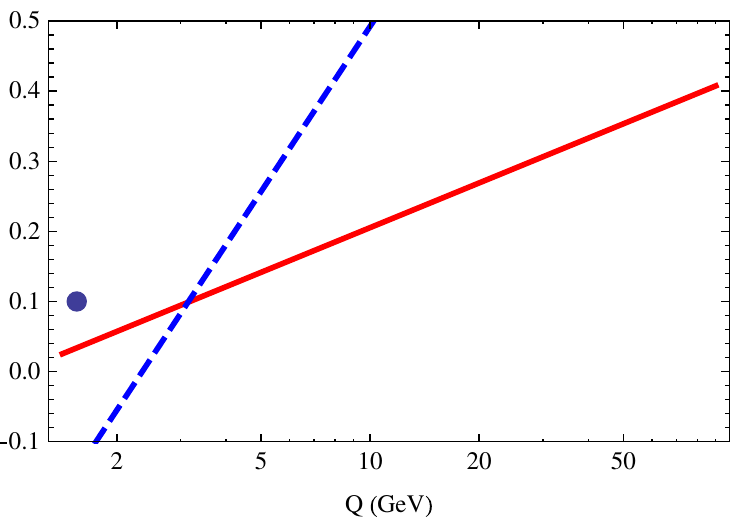}
    \caption{ Coefficient of $-\Tsc{b}^2$ in the exponent in Eq.\
      (\ref{eq:large.b}), from Sun and Yuan \cite{Sun:2013hua}, as a
      function of $Q$ at $x=0.1$.  The blue dashed line is for the
      BLNY fit, and the red solid line for a KN fit with
      $\bmax=\unit[1.5]{GeV^{-1}}$.  The dot represents the value
      needed for SIDIS at HERMES.}
\label{fig:SY}
\end{figure}

At low enough $Q$, the coefficient of $\Tsc{b}^2$ in the exponent
reverses sign, so that the integral diverges at large $\Tsc{b}$
instead of converging.  With the BLNY fit, this reversal of sign
occurs \cite{Sun:2013hua} in a region where there is data and where it
is reasonable to apply TMD factorization.  This is illustrated in
Fig.\ \ref{fig:SY}.  Even with the KN fit using
$\bmax=\unit[1.5]{GeV^{-1}}$, which gives a smaller coefficient of
$\Tsc{b}^2$ in $\tilde{K}$, the evolved exponent is well below what is
needed to fit HERMES data.

\section{Systematic analysis of non-perturbative part of evolution}

I propose the following assertions as a starting point to resolve the
apparent discrepancies and contradictions in the literature,
concerning $\tilde{K}(\Tsc{b})$ at large $\Tsc{b}$:
\begin{itemize}
\item This function (or something equivalent) is needed to implement
  correctly the $Q$ dependence of TMD cross sections.
\item Surely $\Tsc{b}$ above about $\unit[3]{GeV^{-1}}=\unit[0.6]{fm}$
  is in domain of non-perturbative physics, since we know that the
  size of the proton is about $\unit[1]{fm}$.
\item It is difficult to avoid confounding $x$-dependence with
  $Q$-dependence of transverse-momentum distributions.  In measuring
  $\tilde{K}$ one must be careful to analyze data with different $Q$
  at the same value of parton $x$.
\item Fig.\ \ref{fig:SY} strongly suggests that evolution of the shape
  of TMD parton densities slows down at lower $Q$ compared with what
  happens in the data fit by BLNY and KN.
\item Low $Q$ involves larger (more non-perturbative) $\Tsc{b}$ than
  high $Q$.
\end{itemize}

I propose the following general guidelines for modifying current
parameterizations:
\begin{itemize}
\item One should assume that the KN form (with its $\Tsc{b}^2$ form)
  is appropriate only for moderate $\Tsc{b}$, to fit the higher energy
  DY data correctly.  KN is preferred here over BLNY both because it
  gives a better fit, and because its value
  $\bmax=\unit[1.5]{GeV^{-1}}=\unit[0.3]{fm}$ is not excessively
  conservative.
\item As can be seen from Fig.\ \ref{fig:evol}, the data used for the
  KN and BLNY fits constrain $\tilde{K}$ mostly at $\Tsc{b}$ below
  about $\unit[2]{GeV^{-1}}$.
\item But $\tilde{K}(\Tsc{b})$ should flatten out at the higher values
  of $\Tsc{b}$ that are relevant for lower $Q$ experiments (HERMES and
  COMPASS, etc).  
\end{itemize}

\section{$\tilde{K}$ at large $\Tsc{b}$}

\subsection{Basic issues}

\begin{figure}
\centering
  \includegraphics[scale=0.6]{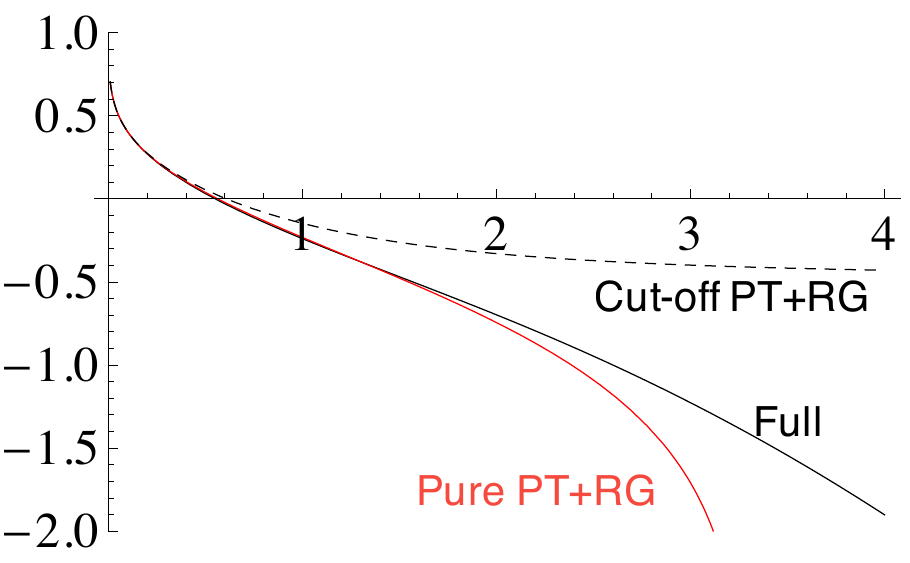}
  \caption{The components of $\tilde{K}$ in (\ref{eq:gK.def}).  It is
    evaluated with the KN parameters for
    $\bmax=\unit[1.5]{GeV^{-1}}=\unit[0.3]{fm}$.  The dashed line is
    the cutoff version $\tilde{K}(\bstarsc,\mu)$, calculated by
    perturbation theory and a standard renormalization-group (RG)
    improvement.  The red solid line is the same thing but with
    $\bmax=\infty$, i.e., it is pure RG-improved perturbation theory.  It
    has a divergence at a finite value $\Tsc{b}$ because of the Landau
    pole in the coupling; perturbation theory is evidently incorrect
    there.  The solid black line gives the full KN result including
    the quadratic fitted $g_K$ function.}
\label{fig:KN-K}
\end{figure}

In this section, I make some remarks on issues about parameterizing
the large $\Tsc{b}$ behavior of $\tilde{K}$.  Within the CSS
$\bstarsc$ prescription, one has
\begin{equation}
  \label{eq:gK.def}
  \tilde{K}(\Tsc{b},\mu)
  = \tilde{K}(\bstarsc,\mu) - g_K(\Tsc{b};\bmax) ,
\end{equation}
where
\begin{equation}
\label{eq:bstar}
  \bstar = \frac{ \T{b} }{ \sqrt{ 1 + \Tsc{b}^2/\bmax^2} }.
\end{equation}
In Eq.\ (\ref{eq:gK.def}), $\tilde{K}(\bstarsc,\mu)$ is intended to be
always perturbative, and all non-perturbative behavior is
parameterized in the function $g_K$.  To illustrate this, Fig.\
\ref{fig:KN-K} shows the decomposition of $\tilde{K}$ with the KN fit.

The fitted value of the $g_K$ function corrects the cut-off
perturbative term, $\tilde{K}(\bstarsc,\mu)$, and brings the result
for the full $\tilde{K}$ back to its RG-improved perturbative value
for $\Tsc{b}$ up to around $\Tsc{b}=\unit[2]{GeV^{-1}}$; only at
higher $\Tsc{b}$ does its curve move away from the diverging pure-PT
line.  One could therefore argue that the fitting has simply
reproduced perturbatively calculable behavior in this extended region,
i.e., up to around $\Tsc{b}=\unit[2]{GeV^{-1}}$, perhaps also that the
$\bstarsc$ method could be improved, and perhaps that
$\bmax=\unit[1.5]{GeV^{-1}}$ is still too conservative.

\subsection{A possible parameterization}

One naive idea is that instead of $\Tsc{b}^2$, one uses the following
parameterization for $g_K$:
\begin{equation}
\label{eq:param1}
    C \left[ \sqrt{\Tsc{b}^2+b_1^2} - \Tsc{b}- b_1 \right] .
\end{equation}
This goes to a constant as $\Tsc{b}\to\infty$.  There are two parameters in
(\ref{eq:param1}).  Better parameterizations can be found.

\subsection{Simple ideas for physics constraints on large $\Tsc{b}$ behavior}

Given the evolution equation (\ref{eq:CSS}), one can characterize
$\tilde{K}(\Tsc{b})$ as quantifying the effects of the emission of
glue for each extra unit of available rapidity, when the energy of an
experiment is increased, at fixed $x$.

So, for extra rapidity range $\Delta y$, let
\begin{itemize}
\item $1-c\Delta y$ = probability of no relevant emission
\item $c\Delta y$ = probability of emitting particle(s)
\item So another possibility for the non-perturbative part of
  $\tilde{K}$ is
    \begin{align}
      \tilde{K}(\Tsc{b})_{\rm NP} 
      &= \text{FT of~} c \left[ - \delta^{(2)}(\T{k}) 
                 + e^{-\Tsc{k}^2/\Tscj{k}{0}^2} / (\pi \Tscj{k}{0}^2)
               \right]
\nonumber\\
      & = c \left[ - 1 + e^{-\Tsc{b}^2\Tscj{k}{0}^2/4} \right].
    \end{align}
\end{itemize}
Here, I have made an ansatz that the transverse-momentum distribution
of non-perturbative particle emission at low transverse momentum is
Gaussian, motivated by commonly used parameterizations.

We get yet another parameterization, now with quadratic behavior at
small $\Tsc{b}$, and a non-infinite limit when $\Tsc{b}\to\infty$.

Perhaps an exponential at large $\Tsc{b}$ instead of a Gaussian would
be better, given known general behavior of correlation functions at
large Euclidean distances, as argued by Schweitzer, Strikman and Weiss
\cite{Schweitzer:2012hh}.

\section{Tool to compare different methods: The $A$ function}

In a separate talk, I proposed a tool that can conveniently be used to
quantitatively compare different methods for TMD factorization in a
scheme-independent way.  It will be described in much more detail in a
forthcoming paper with Ted Rogers.

The motivation arises as follows:
\begin{itemize}

\item The shape change of transverse momentum distribution comes only
  from $\Tsc{b}$-dependence of $\tilde{K}$ in the CSS formalism, or
  from some similar quantity.

\item Generally in any TMD factorization scheme, the cross section can
  be written as a Fourier transformation:
\begin{equation}
  \frac{\diff{\sigma}}{ \diff[4]{q} }
  = \mbox{normalization}
    \times
    \int e^{i\T{q}\cdot \T{b} }
       \widetilde{W}(\Tsc{b},s,x_A,x_B) \diff[2]{\T{b}}
\end{equation}

\item So let us define a scheme-independent function\footnote{The
    function was called $L$ in the talk.  But is now renamed $A$
    because of its essential identity with a function of the same name
    but different arguments in \cite{Collins:1984kg}.}
\begin{align}
  A(\Tsc{b})
  & =
    - \frac{ \partial }{ \partial\ln \Tsc{b}^2 }
      \frac{ \partial }{ \partial\ln Q^2 }
      \ln \tilde{W}(\Tsc{b},Q,x_A,x_B)
\nonumber\\
   & \stackrel{\textrm{CSS}}{=}
    - \frac{ \partial }{ \partial\ln \Tsc{b}^2 }
      \tilde{K}(\Tsc{b},\mu),
\end{align}
where the second line gives its value in the CSS method.  
\item QCD predicts that this function is:
  \begin{itemize}
  \item independent of $Q$, $x_A$, $x_B$,
  \item independent of light-quark flavor,
  \item RG invariant,
  \item perturbatively calculable at small $\Tsc{b}$,
  \item non-perturbative at large $\Tsc{b}$.
  \end{itemize}

\end{itemize}

It will be useful to compare the values of $A(\Tsc{b})$ that
correspond to fits and formula in the different articles on the
subject of TMD factorization and evolution.  The values of parameters
where discrepancies occur can be used as a diagnostic: To show which
experimental data will be most incisive in arbitrating the correctness
of different treatments, and to diagnose which treatments are in
disagreement with QCD and whether the disagreements are significant.

\section{Concluding remarks}

\begin{itemize}
\item Surely we need non-perturbative contributions to TMD
  factorization.  The values of $\Tsc{b}$ that are important in the
  Gaussian parameterizations of TMD densities are in a region not far
  from the proton size.  Everybody agrees that some parameterization
  of the non-perturbative properties of TMD densities is needed to
  describe data at low enough transverse momentum (and hence at large
  $\Tsc{b}$). 
\item Therefore one must also understand their evolution in this same
  non-perturbative region of large $\Tsc{b}$.
\item According to established theorems, evolution of TMD functions is
  governed by a single universal function, $\tilde{K}$ or some
  equivalent.
\item Extrapolation of earlier DY fits to use them at the values of
  $\Tsc{b}$ relevant for lower energy SIDIS is incorrect.
\item It is essential to use better parameterizations of $\tilde{K}$
  so that at large $\Tsc{b}$ its functional form flattens.  The
  parameterizations should be such that they retain compatibility with
  the evolution measured in Drell-Yan experiments, where substantially
  smaller values of $\Tsc{b}$ are important compared those needed for
  the data from the HERMES and COMPASS experiments.
\item Physical and phenomenological arguments were given in support of
  these assertions.
\item It is necessary to redo global fits with better
  parameterizations, and a clear sense of which data are relevant for
  which regions of transverse position $\Tsc{b}$.
\item In testing and measuring TMD evolution it is essential to ensure
  that the data being compared are at fixed $x$ with different $Q$.
\item A large coefficient for the $\Tsc{b}^2$ term in $\tilde{K}$ (and
  $g_K$) at large $\Tsc{b}$ causes substantial dilution of the Sivers
  asymmetry, etc, at large $Q$, thereby requiring greater sensitivity
  in future higher-energy experiments.  Getting improved understanding
  and measurements of the non-perturbative part of TMD evolution is
  important to planning these future experiments.
\end{itemize}


\section*{Acknowledgments}

This work was supported by the U.S. Department of Energy.  I thank
many colleagues for discussions, notably Ted Rogers and Ahmad Idilbi.

\bibliography{jcc}

\end{document}